\documentstyle[preprint,aps]{revtex}
\begin{document}
\draft
%


\title {\Large \bf
Inter-Condensate Tunneling in Bose-Einstein 
Condensates with Feshbach Resonances}
\author{ 
Eddy Timmermans$^{1}$, Paolo Tommasini$^{1}$, Robin C\^{o}t\'{e}$^{1}$, Mahir Hussein$^{2,1}$\\
and Arthur Kerman$^{3}$\\
$^{1}$Institute for Atomic and Molecular Physics \\
Harvard-Smithsonian Center for Astrophysics 
60 Garden Street, Cambridge, MA 02138 \\
$^2$ Instituto de F\'{i}sica, Universidade de S\~{a}o Paulo, C.P. 
66318, \\
CEP 05315-970 S\~{a}o Paulo, Brazil \\
$^{3}$ Center for Theoretical Physics, Laboratory for Nuclear Science and \\
Department of Physics, Massachusetts Institute of Technology\\
Cambridge, MA 02139}
\date{\today}
\maketitle
\begin{abstract}

	Recently, effects of Feshbach resonances in atom-atom
interactions were observed by varying the external magnetic
field of an atomic Bose-Einstein condensate (BEC).
We point out that the quasi-bound molecules created in the
intermediate state of the resonance can form a second, molecular
condensate.  The many-body state of the system is then a hybrid
atomic/molecular condensate with inter-condensate tunneling of atom pairs.
A sudden variation of the magnetic field results in oscillations
of the number of atoms and molecules in their respective condensates,
providing a signature of this novel type of quantum tunneling.

\end{abstract}



	Recently, Ketterle's group reported evidence for magnetic-field-induced
Feshbach resonances \cite{b1} in the interparticle interactions of atomic-trap
Bose-Einstein condensates \cite{b21}-\cite{b23}.  
Near resonance, the two-body interactions
are characterized by an effective interaction strength that can be tuned by 
varying the magnetic field, suggesting the construction of condensates
with tunable interaction strength \cite{b3}.  
We present a theoretical interpretation 
of the condensate Feshbach resonances that predicts the formation of a
molecular condensate that is coupled to the atomic condensate by quantum 
tunneling of atom pairs.

\section{The Binary Collision Resonance}

	In an external magnetic field, the hyperfine induced spin
flip of one atom in the presence of a second atom
can bind the interacting atoms by bringing
them to a spin configuration of higher energy.  A resonance occurs
if the potential of the bound channel supports an eigenstate that
is nearly degenerate with the incident atoms \cite{b41}-\cite{b45}.

	The detuning of the resonance, $\epsilon$, is the energy difference
of the quasi-bound molecular state and the continuum of the
incident atoms, pictured in Fig.(1).  The energy difference $\Delta$
between the continua of the initial and quasi-bound channels, 
is the energy gain
of a single spin flipped atom in the magnetic field $B$.  
When $B$ takes on its resonant value, $B_{0}$, 
the energy of the quasi-bound
molecule lines up with the continuum of the incident atoms
so that $\Delta$ equals the
binding energy of the bound state, $\Delta = E_{b}$.  
Near resonance, $\Delta \approx E_{b} + (\partial \Delta / \partial B)
\times
[B - B_{0}]$, and one controls the detuning by varying the magnetic field, 
$\epsilon = \Delta - E_{b} \approx (\partial \Delta / \partial B) 
\times [B - B_{0}]$.

	In the many-body Hamiltonian, 
the spin flips to quasi-bound molecules are described by
\begin{equation}
H_{FR} = \alpha \int d^{3} r \; \hat{\psi}_{m}^{\dagger} ({\bf r})
\hat{\psi}_{a} ({\bf r}) \hat{\psi}_{a} ({\bf r}) \; + h. c. \; \; ,
\label{e:e1}
\end{equation}
where $\hat{\psi}_{m} , \hat{\psi}_{m}^{\dagger}$ (
$\hat{\psi}_{a}, \hat{\psi}_{a}^{\dagger}$) are the annihilation
and creation field operators of the molecules (atoms).
The $\alpha$-parameter in Eq.(\ref{e:e1}) is the transition matrix
element proportional to the overlap of the molecular continuum and bound state
wave functions.

	The $H_{FR}$ interaction is responsible for the Feshbach
resonance and gives a dispersive
contribution to the effective interaction strength.  
To find the off-resonant detuning dependence
we determine the energy
shift of two atoms confined to a volume $\Omega$.
If the initial state $|{\rm ini} \rangle$ of energy
$E_{{\rm ini}}$ has both atoms in the same momentum state, then
the interatomic interaction,
$[\lambda_{a}/2] \int d^{3} r \hat{\psi}_{a}^{\dagger}({\bf r})
\hat{\psi}_{a}^{\dagger}({\bf r}) 
\hat{\psi}_{a}({\bf r}) \hat{\psi}_{a}({\bf r})$, contributes a shift
$\Delta E_{a} = \lambda_{a}/\Omega$ in first order perturbation. 
If we denote the quasi-bound state 
of energy $E_{{\rm int}}$ by $|{\rm int}\rangle$, then the second order
shift $\Delta E_{FR}$ due to $H_{FR}$ is given by $\Delta E_{FR}$ =
$|\langle {\rm int} | H_{FR} | {\rm ini} \rangle |^{2}
/ [ E_{{\rm ini}} - E_{\rm{int}} ]$ = $- [2 \alpha^{2} /\epsilon]/\Omega$.
Thus, $\Delta E = \Delta E_{a} + \Delta E_{FR} = \lambda_{{\rm eff}}
/\Omega$, where the effective interaction strength is
equal to $\lambda_{{\rm eff}} = \lambda_{a} - 2 \alpha^{2}/\epsilon$.

\section{Feshbach-Resonance Interactions in the Condensate}

	From momentum conservation, it follows that all quasi-bound molecules
created from condensate atoms of vanishing momentum also occupy the same
zero-momentum center-of-mass state, giving a first indication 
that the quasi-bound molecules form a condensate.
A more insightful argument follows from the Heisenberg equations
$i \hbar \hat{\dot{\psi}}_{a} = 
[ \hat{\psi}_{a}, \hat{H} ]$, $i \hbar \hat{\dot{\psi}}_{m} =
[ \hat{\psi}_{m}, \hat{H} ]$, where $\hat{H}$ is the Hamiltonian.
The expectation value of these operator equations
gives equations of motion for the condensate fields, $\phi_{m}
= \langle \hat{\psi}_{m} \rangle$ and $\phi_{a} = \langle \hat{\psi}_{a}
\rangle $.  
In the mean-field approximation (e.g. $\langle \hat{\psi}_{a} ({\bf r}) \hat{\psi}_{a} ({\bf r}) \rangle 
\approx \phi_{a}^{2} ({\bf r})$) \cite{b4a}, this procedure yields two coupled 
equations in place of the
usual single condensate Gross-Pitaevskii equation \cite{b51}-\cite{b52}:
\begin{eqnarray}
i \hbar \dot{\phi}_{m} &=&
[ - \frac{\hbar^{2} \nabla^{2} }{4 M} + \epsilon + \lambda_{m} n_{m}
+ \lambda n_{a} ] \phi_{m} + \alpha \phi_{a}^{2} 
\nonumber \\
i \hbar \dot{\phi}_{a} &=&
[ - \frac{\hbar^{2} \nabla^{2}}{2 M} + \lambda_{a} n_{a} 
+ \lambda n_{m} ] \phi_{a} + 2 \alpha \phi_{a}^{\ast} \phi_{m}
\; \; ,
\label{e:e2}
\end{eqnarray}
where $M$ denotes the mass of a single atom,
$n_{m}$ and $n_{a}$ represent the condensate densities, 
$n_{m} = |\phi_{m}|^{2}$ and $n_{a} = |\phi_{a}|^{2}$, and 
$\lambda_{m},
\lambda_{a}$ and $\lambda$ represent the strength of the molecule-molecule,
atom-atom and molecule-atom interactions.  The $\alpha$-terms that
couple the equations
describe tunneling of pairs of atoms between
the $\phi_{m}$ and $\phi_{a}$-fields.  In particular, $\phi_{m}$ acquires
a source term, $\alpha \phi_{a}^{2}$, so that
its value cannot remain zero if $\phi_{a}$ is finite: the tunneling
creates a molecular condensate that is coherent with the atomic condensate.

\section{ Off-resonant Statics}

	Provided the condensate lifetime exceeds the time required
by the system to reach its equilibrium, a discussion of
the `ground state' of the many-body system is meaningful.
The equations that describe the static system are similar to Eqs.(2)
with the time derivatives replaced by the chemical potential:
$i \hbar \dot{\phi}_{a} \rightarrow \mu \phi_{a}$ and 
$i \hbar \dot{\phi}_{m} \rightarrow 2 \mu \phi_{m}$, where the chemical
potential of the molecules is twice the chemical potential of the atoms,
in accordance with the condition for chemical equilibrium.

	Of particular interest is the off-resonant ($\epsilon > 0$)
limit: although
$\epsilon$ is tuned close enough to observe resonance effects, its value
still exceeds the kinetic energy and interaction energies ($\lambda_{m} n_{m},
\lambda n_{m}, \lambda_{a} n_{a}$ and $\lambda n_{a}$).  As a consequence,
$\epsilon >> \mu$, and $n_{m}<< n_{a}$.  The $\phi_{m}$-equation 
yields $\phi_{m} \approx
- \alpha n_{a}/\epsilon$ (where we take $\phi_{a}$ to be real) and
the $\phi_{a}$-equation yields a Gross-Pitaevskii equation with the 
effective binary collision interaction strength:
\begin{equation}
\mu \phi_{a} = \left[ -\frac{ \hbar^{2} \nabla^{2}}{2M} + (\lambda_{a} - 
\frac{2 \alpha^{2}}{\epsilon} ) n_{a} \right] \phi_{a} \; \; .
\label{e:e3}
\end{equation}
Note that equation (\ref{e:e3}) by itself does not describe the appearance
of the small molecular condensate, a feature that could have important
applications \cite{b5a}.

	In the off-resonant limit we can also understand the increase
in loss rate that served as a signal to detect the Feshbach resonance.
As the fraction of molecules remains low,
we can assume that the atomic condensate decays mostly due to inelastic
atom-atom collisions $\dot{n}_{a} = 
- c_{aa} n_{a}^{2}$, and the molecular condensate due
to molecule-atom collisions, $\dot{n}_{m} = - c_{ma} n_{a} n_{m} 
\approx - n_{a}^{3} c_{ma} [\alpha/\epsilon]^{2}$, where $c_{aa}$
and $c_{mm}$ represent the corresponding rate coefficients.  If we count
each $a$-atom as one and each $m$-molecule as two condensate particles, the
condensate particle density is $n = n_{a} + 2 n_{m}$, 
and the condensate loss is described by
$\dot{n} = \dot{n}_{a} + 2 \dot{n}_{m} \approx - n^{2}
( c_{aa} + 2 c_{am} n [\alpha/\epsilon]^{2} )$.

	The observed Feshbach resonances create quasi-bound molecules
of high vibrational quantum number.
The single particle lifetime of such loosely bound molecules,
limited due to collisions with atoms and other molecules, is substantially
lower than the lifetime of the individual atoms.  
Estimates for the alkali-dimer rate coefficients 
give $10^{-11}$ -- $10^{-9}$ $cm^{3}/sec$ \cite{b6}, compared
to $c_{aa} \sim 10^{-14} cm^{3} sec^{-1}$.  Thus, a purely
molecular condensate of density $10^{14} cm^{-3}$ does not live longer than
$10^{-3}$ seconds.  In contrast, the molecular condensate in an off-resonant
BEC survives much longer by
compensating for the loss of molecules with atom pairs that tunnel
in from the atomic condensate.

\section{Dynamics}

	The dynamics of a condensate in a time varying magnetic field, 
giving
a time-dependent detuning $\epsilon (t)$, is particularly interesting. 
A proper description requires an adjustment of Eqs.(2) to account 
for condensate loss.

	A formal treatment of the loss includes the 
channels of all chemical reactions that remove 
particles from the condensate.  The two-body collision
channels can be eliminated in perturbation theory,
modifying the equations of motion (Eqs.(\ref{e:e2}))
in a simple and predictable way: the interaction strengths become
`absorptive' with an imaginary part that
determines the loss rates.  In the $\phi_{m}$-equation, for example,
$\lambda \rightarrow \lambda - i \hbar c_{ma}/2$ and
$\lambda_{m} \rightarrow \lambda_{m} - i \hbar c_{mm}/2$, where $c_{mm}$
is the rate coefficient accounting for molecule-molecule collisions.

	In the off-resonant limit, the atomic density
variations can be neglected and the atomic
condensate serves as a reservoir of atoms. 
The appropriate approximations to the
$\phi_{m}$-equation of Eqs.(\ref{e:e2}), gives 
\begin{equation}
i \hbar \dot{\phi}_{m} = \left[ \epsilon (t) + \lambda n_{a} 
- i \hbar \frac{\gamma_{m}}{2} \right] \phi_{m} + \alpha \phi_{a}^{2} (t)
\; \; ,
\label{e:e4}
\end{equation}
where $\gamma_{m}$ is the molecule loss-rate: $\gamma_{m} = c_{ma} n_{a}$.
In the same off-resonant limit, 
$\phi_{a} \approx \sqrt{n_{a}} \exp [-i \lambda_{a} n_{a} t / \hbar]$.

	For the special case that the detuning undergoes a sudden shift
to a value $\epsilon_{f}$ and remains constant thereafter, we find 
\begin{eqnarray}
\phi_{m}(t) &=& \phi_{\infty} \exp \left[ 
\frac{-i \; t}{\hbar} (2 \lambda_{a} n_{a}) \right] +
\nonumber \\
&& \; \; \; [ \phi_{0} - \phi_{\infty} ] \times
\exp \left[ \frac{- i\; t}{\hbar} \left( \epsilon_{f} + \lambda n_{a}
\right) \right] \; 
\exp(\frac{-\gamma_{m} t}{2} ) 
  \; \; ,
\label{e:e5}
\end{eqnarray}
where $\phi_{0}$ is the initial field, $\phi_{0} = \phi_{m}(t=0)$, and
$\phi_{\infty}$ is the static off-resonant field value,
$\phi_{\infty} = - \alpha n_{a} / [ \epsilon_{f} + (\lambda - 2 \lambda_{a})
n_{a} - i \hbar \gamma_{m}/2 ] \approx - \alpha n_{a} / \epsilon_{f}$.
Note that the molecular density has an oscillating contribution
$\sim 2 |(\phi_{0} - \phi_{\infty}) \phi_{\infty}^{\ast}| \;
\cos ( [\epsilon_{f} + \lambda n_{a} - 2 \lambda_{a} n_{a} ) ] t/\hbar) 
\exp(-\gamma_{m} t/2)$.
The molecules that appear and disappear during the oscillations tunnel in 
and out from the atomic condensate so that the atomic density has a
contribution that oscillates out of phase with twice the amplitude.
The oscillation is a genuine quantum tunneling effect and stems 
from the interference of the propagating initial
field amplitude with the propagating amplitude of atom-pair tunneling. 
In general, the oscillations appear if the detuning was changed at a 
rate $|\dot{\epsilon}|$ that exceeds $|\epsilon \gamma_{m}|$, or
$|\dot{B}/[B-B_{0}]| >> \gamma_{m}$.  In the opposite limit,
$|\dot{\epsilon}/\epsilon| < \gamma_{m}$, the system adiabatically follows
its ground state.
Experimentally, the oscillation can be observed, for example, by  
illuminating the BEC with light that is resonant with a transition
of the quasi-bound molecule.  The intensity of the image will be modulated
at the frequency of the quantum tunneling oscillations.

	Near resonance, the nonlinear density variations cannot
be neglected, but the dynamical behavior remains qualitatively
similar: a sudden shift of detuning results in out-of-phase oscillations
of the atomic and molecular condensate densities, as shown in Fig.(2).
The oscillations are damped and
the condensate density decays on a longer time scale.

\section{Conclusions}

	In an atomic BEC, the molecules created in the intermediate state of a 
Feshbach resonance form a second condensate that is coherent with the
atomic condensate field.  Pairs of atoms tunnel between the condensates.
We have discussed a clear signature of this novel type of quantum tunneling:
a sudden change in the external magnetic field results in oscillations
of the number of atoms and molecules in their respective condensates.

We thank P. Milonni and A. Dalgarno for interesting discussions.
The work of E.T. and R. C. was supported by the
NSF through a grant for the
Institute for Atomic and Molecular Physics at Harvard University
and Smithsonian Astrophysical Observatory,  of M. H.
in part by the Brazilian agencies CNPq and Fapesp.
and of A. K. in part by the U.S. Department of Energy
(D.O.E.) under cooperative research agreement DEFC02-94ER418.

\newpage
\centerline{\bf Figure Captions}
\noindent
\underline{Fig.1 }:
	Schematic representation of the molecular potentials of
the incident and intermediate state channels .  The
energy difference of the continuum levels, $\Delta$, is the sum of the
binding energy $E_{b}$ of the quasi-bound state
and the `detuning' $\epsilon$.

\vskip 0.15in

\noindent
\underline{Fig.2 }:
	Plot of the particle
densities: the total condensate density,
 $n = n_{a}  + 2 n_{m}$, in full line,   
the atomic density $n_{a}$ in dashed line and 
the molecular density $n_{m}$ in dash-dotted
line.  The oscillatory behavior of the atomic and molecular densities 
is a signature of quantum tunneling.  The calculation
is for a homogeneous BEC that was initially in equilibrium
at density $n=10^{14} cm^{-3}$ when the detuning experienced
a sudden shift from  $\epsilon = 50 \lambda n$ to
$\epsilon = 2 \lambda n$.  The order of magnitude of the interaction parameters,
$\lambda n = \lambda_{m} n = \lambda_{a} n = \alpha \sqrt{2n} =
10^{5}$ Hz, and of the decay parameters,
$c_{ma} = c_{mm} = 5 \times 10^{-10} cm^{3} sec^{-1}$ (while
neglecting the atomic decay) are realistic.


\begin{thebibliography}{99}
\bibitem{b4a} We assume a dilute zero temperature system so that
we can neglect condensate depletion.
\end{thebibliography}
\end{document}